\documentclass[twocolumn]{aastex63}

\received{xx}
\revised{xx}
\accepted{February 1, 2023}
\submitjournal{ApJ}

\shorttitle{SMC UV catalogue}
\shortauthors{A. Devaraj et al.}

\watermark{}
\setwatermarkfontsize{15pt}

\graphicspath{{./}{figures/}}

\usepackage{xcolor}
\usepackage{longtable}

\begin{document}

\title{UVIT observations of the Small Magellanic Cloud: Point source catalogue}

\correspondingauthor{A. Devaraj}
\email{ashidevaraj@gmail.com}

\author[0000-0001-5933-058X]{A. Devaraj}
\affiliation{Indian Institute of Astrophysics, Bangalore, Karnataka, 560034, India}
\affiliation{Department of Physics and Electronics, CHRIST (Deemed to be University), Bangalore, Karnataka, 560029, India}

\author[0000-0003-1409-1903]{P. Joseph}
\affiliation{Indian Institute of Astrophysics, Bangalore, Karnataka, 560034, India}
\affiliation{Department of Physics and Electronics, CHRIST (Deemed to be University), Bangalore, Karnataka, 560029, India}

\author[0000-0002-4998-1861]{C. S. Stalin}
\affiliation{Indian Institute of Astrophysics, Bangalore, Karnataka, 560034, India}

\author{S. N. Tandon}
\affiliation{Inter-University Center for Astronomy and Astrophysics, Pune, Maharashtra, 411007, India}

\author{S. K. Ghosh}
\affiliation{Tata Institute of Fundamental Research, Mumbai, Maharashtra, 400005, India}

\begin{abstract}

Three fields in the outskirts of the  Small Magellanic Cloud were observed by the Ultra-Violet Imaging Telescope (UVIT) on board {\it AstroSat}, during 31 December 2017 and 01 January 2018.  The observations were carried out on a total of seven filters, three in the far ultra-violet (FUV; 1300$-$1800 \AA) band and four in the near ultra-violet (NUV; 2000$-$3000 \AA) band. We carried out photometry of these observations that have a spatial resolution better than 1.5$^{\prime\prime}$. We present here the first results of this work, which is a matched catalogue of 11,241 sources detected in three  FUV and four NUV wavelengths. We make the catalogue available online, which would be of use to the astronomical community to address a wide variety of astrophysical problems. We provide an expression to estimate the total count rate in the full point spread function of UVIT that also incorporate the effect of saturation.
\end{abstract}

\keywords{Ultraviolet astronomy (1736) -- Ultraviolet telescopes (1743);
Ultraviolet photometry (1740)}


\section{Introduction} \label{sec:intro}
Small Magellanic Cloud (SMC) is one of the closest (D=61.9 $\pm$ 0.6 kpc;
\citealt{2015AJ....149..179D}) star forming galaxies to our Galaxy \citep{2005MNRAS.357..304H}. It has a low metallicity with Z = 0.005 \citep{1984IAUS..108..353D} and low foreground extinction of E(B$-$V) = 0.02 mag \citep{1982ApJ...255...70H}. The 2175 \AA ~bump is absent in SMC which could be due to the dust in SMC being different from either the Milky Way or the Large Magellanic Cloud (LMC) and moreover, this has been attributed to the lack of carbonaceous dust \citep{2001ApJ...548..296W}. SMC has been surveyed at various wavebands such as the near-infrared by the Two  Micron All Sky Survey (2MASS; \citealt{2003AJ....126.1090C}) in the mid and far-infrared by {\it Spitzer} \citep{2011AJ....142..102G} and in the optical \citep{2002ApJS..141...81M}. These observations indicate that SMC can be a unique laboratory to investigate stellar evolution and interstellar matter at low metallicity environment. SMC has also been targeted for observations in the X-ray band for studies on the X-ray binary population in low metallicity conditions \citep{2019ApJ...884....2L}. In spite of the various multi-wavelength observations available on SMC, the effect of its low metallicity appears most significant in the ultra-violet band  (UV;\citealt{1997AJ....113.1011C}). For example, as the spectral energy distribution of hot stars peaks at short wavelengths, far ultra-violet (FUV) observations are important to determine the temperature of those hot stars compared to optical or infrared photometry. Though observations of SMC in the UV bands is highly important, a complete census of point sources (at a resolution similar to that available in the optical and near-infrared) is missing. SMC has been observed in the past by the Hubble Space Telescope (HST), the ultra-violet imaging telescope (UIT) flown on Space shuttle during Astro-1, Astro-2 missions \citep{1997AJ....113.1011C,1994ApJ...430L.117C} and {\it Swift}/UVOT \citep{2017MNRAS.466.4540H}. The region of SMC to a large extent has been covered by the Galaxy Evolution Explorer (GALEX; \citealt{2014AdSpR..53..939S}) , though only in the near ultra-violet (NUV) band (1771$-$2831 \AA) with a spatial resolution of around 5 arcsec. Though UIT observations are at a better spatial resolution (3 arcsec) than GALEX, such observations both in FUV and NUV are available only for limited regions of SMC. There is thus a need to improve the coverage and depth of the observations of SMC in both the FUV and NUV bands.

The Ultra-Violet Imaging Telescope (UVIT) on board India's multi-wavelength astronomy satellite called {\it AstroSat} \citep{2006AdSpR..38.2989A} was launched on 28 September 2015. UVIT observes simultaneously in the FUV (1300$-$1800 \AA)  and NUV (2000$-$3000 \AA) bands \citep{2020AJ....159..158T} and provides better resolution images than GALEX and  UIT. The main motivation of this work is to provide a point source catalogue for about 1/4 square degree of SMC field in multiple narrower filters in FUV and NUV at a resolution comparable to typical ground based observations in the visible band. The observations and data reduction are described in Section 2, the generation of the point source catalogue is given in Section 3 followed by the summary in the final Section.

\begin{figure*}[ht!]
\plotone{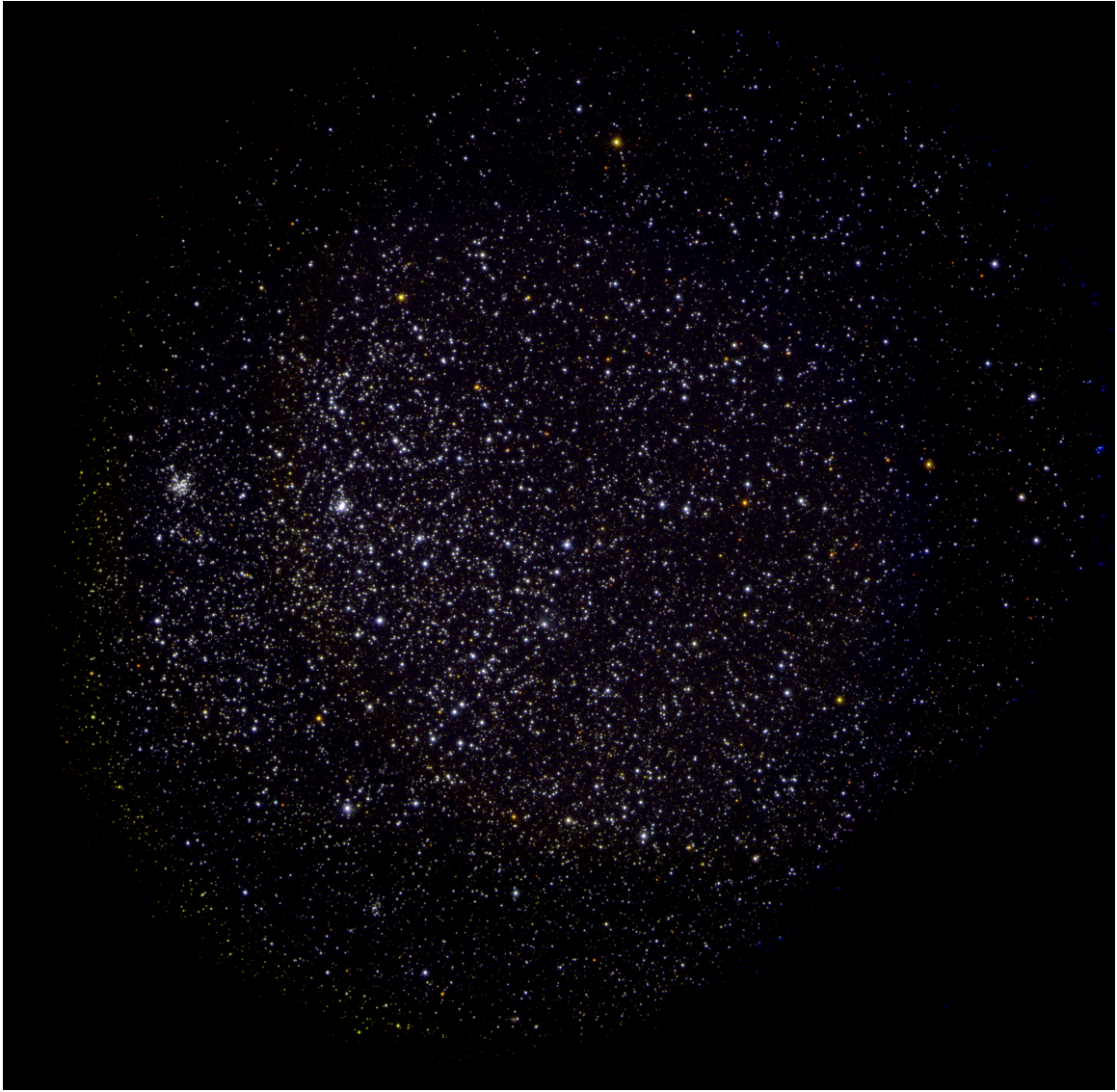}
\caption{An RGB mosaic image of SMC-1, SMC-2, and SMC-3 fields. Here, red, green, and blue refer to the observations made in N263M, N245M, and F154W filters respectively. \label{figure-1}}
\end{figure*}


\section{Observations and Reductions} \label{sec:obs}
The observations used in this work were taken by UVIT. UVIT consists of two 38 cm telescopes, one telescope for FUV and the second telescope for both NUV and VIS (3200$-$5500 \AA) wavelengths. It has a circular field of view of 28 arcmin diameter and provides images with spatial resolution better than 1.5$^{\prime\prime}$. It also has several filters in each of the channels \citep{2020AJ....159..158T}. The VIS channel is primarily used for tracking the aspect of the telescope during observation and applying offline corrections for the spacecraft drift and other disturbances.

Three SMC fields were observed by UVIT during 31 December 2017 and 01 January 2018 (see Table \ref{table-1} for details). These exposures were used primarily to find flat field variations across the 20$'$ field of view, for all the detector-filter combinations in NUV and FUV. The results of these are given in \cite{2020AJ....159..158T}.

The first field (SMC-1) was selected far away from  the central part of SMC so as to avoid the bright central regions of SMC and  centered at $\alpha_{2000}$ = 01:09:46 and $\delta_{2000}$ = $-$71:20:30.0. The second (SMC-2) and third (SMC-3) fields were pointed to by applying a shift of $\sim$6 arcmin in  orthogonal directions. A total of seven filters were used for the observations. Figure \ref{figure-1} shows the RGB image of the three fields. The details of the observations are given in Table \ref{table-1}. The effective wavelength and the bandwidth of the filters used in this work are given in Table \ref{table-2}. More details such as the effective areas of these filters can be found in \cite{2017AJ....154..128T,2020AJ....159..158T}.

\begin{figure*}
\begin{center}
\vbox{
      \includegraphics[scale=0.41]{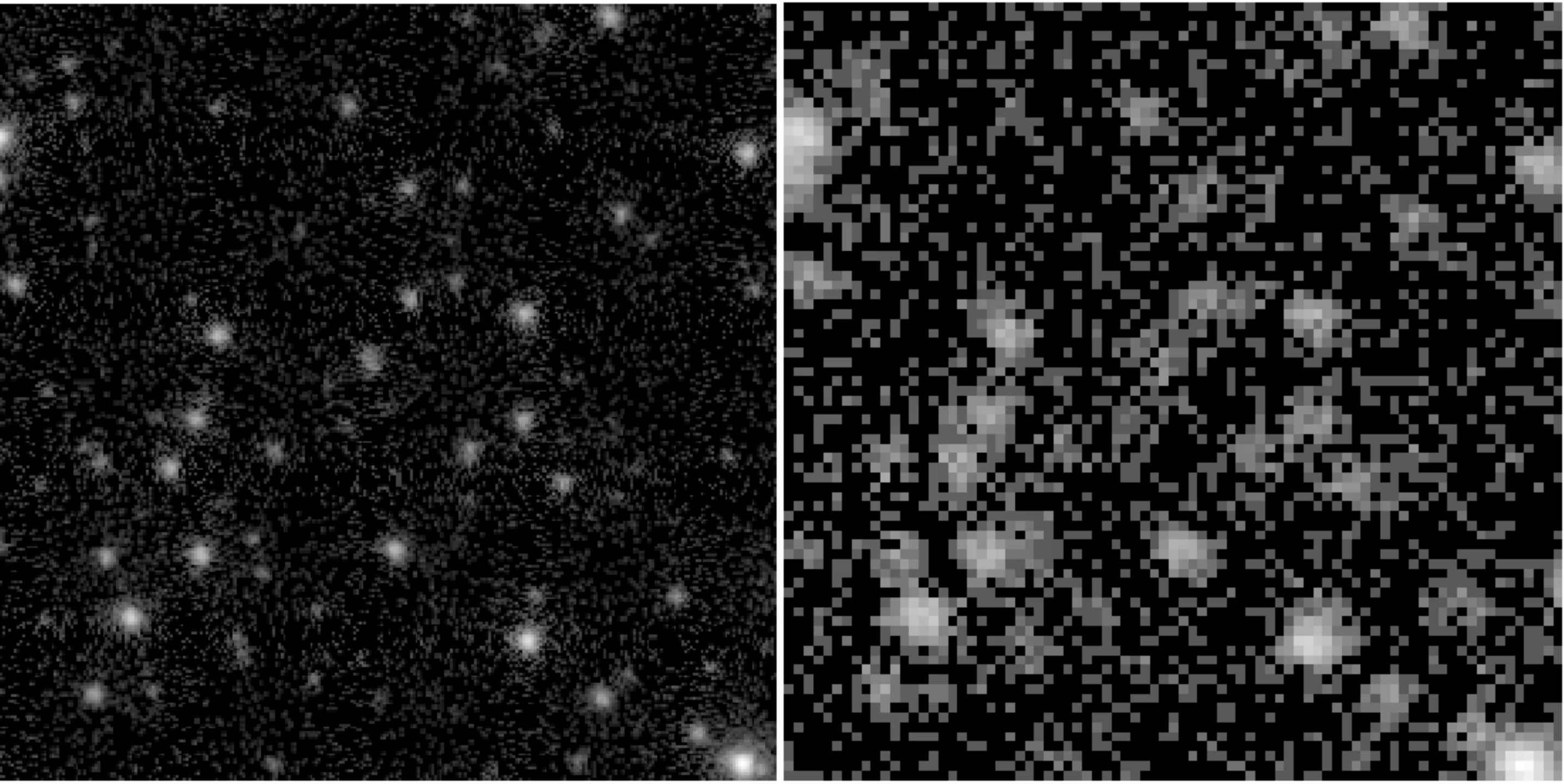}

      \includegraphics[scale=0.41]{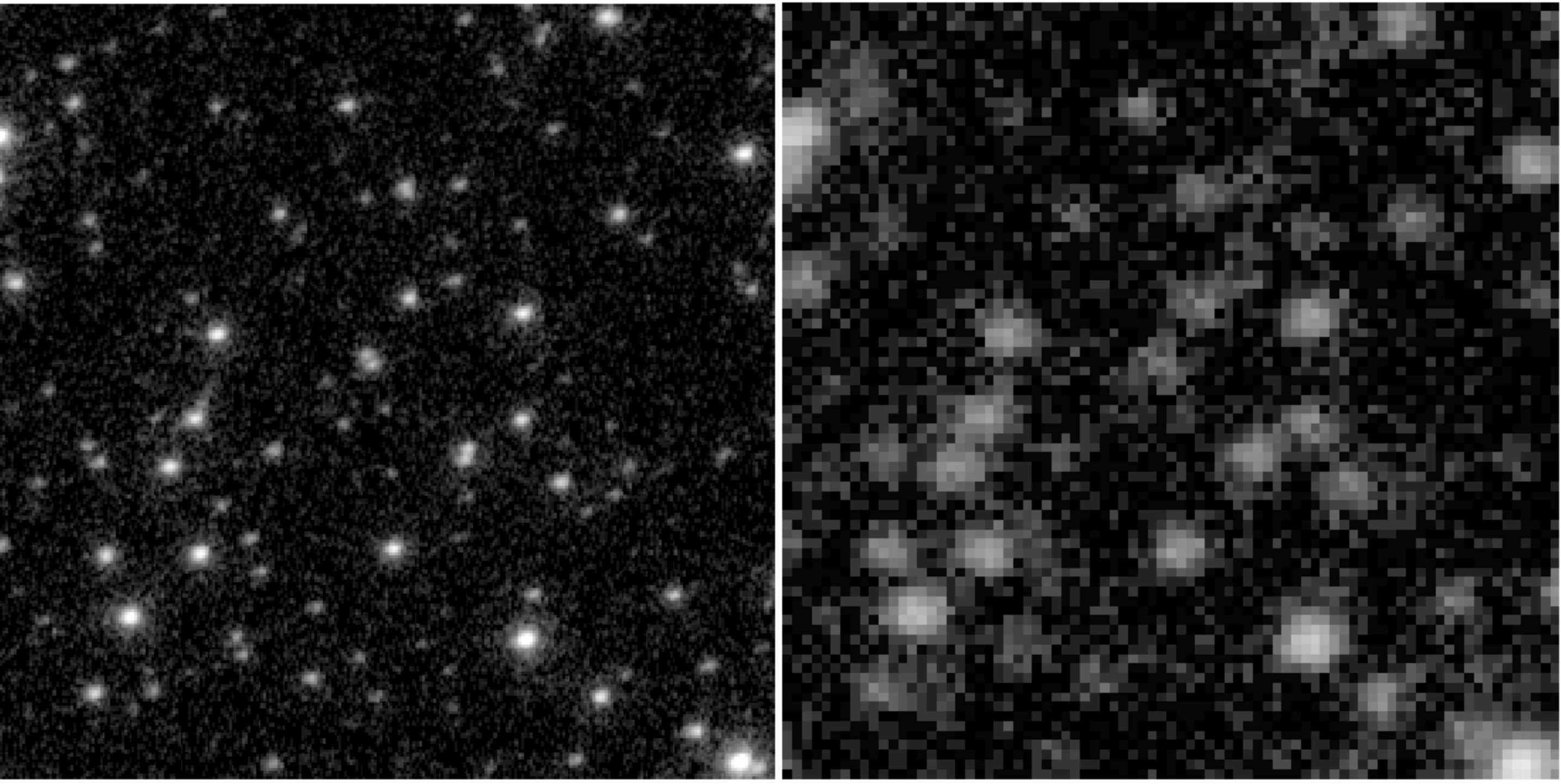}
     }
\end{center}
\caption{A 2 arcmin $\times$ 2 arcmin region of SMC-1 centered at
$\alpha$ = 01:09:46.0, $\delta$ = $-$71:20:30.0. The top left panel shows the UVIT FUV image in F154W filter, while the top right panel is the GALEX FUV image. The bottom panels show the image of the same region in UVIT NUV in N245M filter (left panel) and in GALEX NUV (right panel)}. 
\label{figure-2}
\end{figure*}

\begin{figure}
\centering
\vbox{
\hspace*{-0.6cm}\includegraphics[scale=0.22]{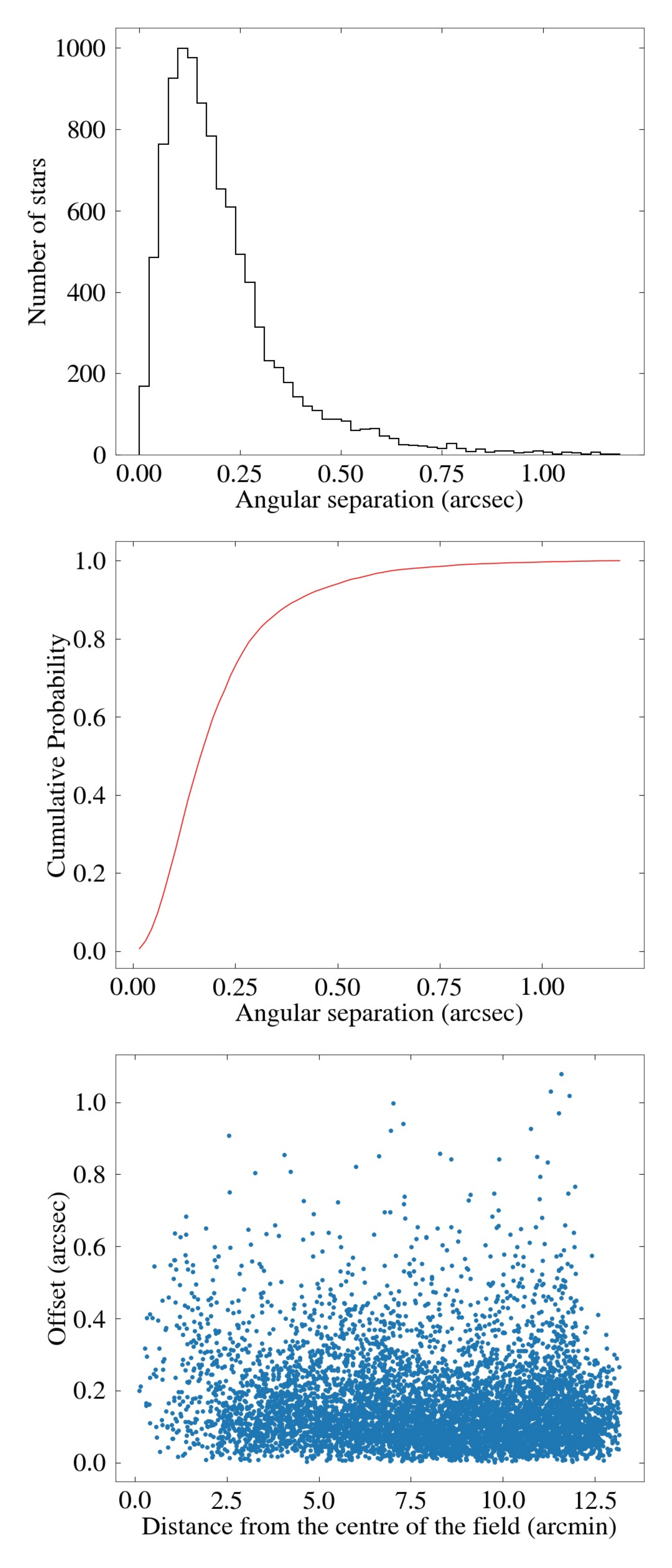}
}
\caption{The distribution of the angular separation between the sources in the UVIT SMC field cross-matched with {\it Gaia} catalogue is given in top panel and the cumulative distribution function of the angular separation is shown in the middle panel. The offset in angular separation between UVIT and Gaia sources as a function of angular distance from the center of SMC-1 field is shown in the bottom panel.}
\label{figure-3}
\end{figure}

The observed images of SMC were reduced using the UVIT L2 pipeline version 6.3 \citep{Ghosh2021, 2022arXiv220307693G}.  This pipeline corrects the observations for geometric distortion, flat field as well as the spacecraft drift. The spacecraft drift was obtained by tracking stars in the field of the VIS channel observations which was then applied to the data acquired in the FUV and NUV channels. The pipeline also performs astrometry of the final images using UV and optical catalogues. The final output of the L2 pipeline is a set of science ready images that includes orbit-wise images  as well as  combined images, wherein the orbit wise images (matched filter wise) are stacked to get better S/N. The central 2 $\times$ 2 arcmin region of SMC-1 observed by UVIT and GALEX is shown in Fig. \ref{figure-2} for comparison of resolution. From Fig. \ref{figure-2}, it is evident that the UVIT image has better resolution than GALEX thereby enabling photometry of more point sources than that possible on the image from GALEX. The astrometry of the final combined images returned by the L2 pipeline is better than few arcsecs, however, to improve the astrometry of UVIT images, we proceeded as follows. Using Aladdin\footnote{https://aladin.u-strasbg.fr/AladinLite/},  we displayed the GALEX image of each of the SMC fields and overlaid the \textit{Gaia} catalogue. From this we visually identified 10 isolated stars in each of the SMC fields spread over the UVIT field of view. For those 10 stars (in each field) we found the (x,y) centroid positions in UVIT images and their corresponding ($\alpha$,$\delta$) from {\it Gaia} Data Release 2 \citep{2018A&A...616A..14G}.The selected stars have negligible proper motion ($<$ 0.88 milli arcsec / year). These information were used in CCMAP routine in {\sc IRAF}\footnote{IRAF stands for Image Reduction and Analysis Facility} to arrive at the transformation between (x,y) and ($\alpha$,$\delta$) that also includes rotation. This transformation was then applied to the UVIT images using CCSETWCS in IRAF, to arrive at the UVIT images with new WCS (World Coordinate System). For doing this, the image taken in the FUV band F154W was considered as the reference and all the other images (both in FUV and NUV) were aligned to it. The WCS of all the images were further refined by an iterative process to minimize the angular separation between {\it Gaia} and {UVIT} coordinates. The distribution of the angular separation between the UVIT ($\alpha$, $\delta$) values and the matched sources with respect to the {\it Gaia} ($\alpha$, $\delta$) values are given in the top panel of Fig. \ref{figure-3}. The cumulative distribution of the same is given in the middle panel of Fig. \ref{figure-3}. It shows that about 90\% of the sources match within 0.4 arcsec. We note that for a separation of about 0.4 arcsec, which includes more than 90\% of the sources, the probability of chance matching with a Gaia source is $\sim$1/250. For the highest separation listed, the probability increases to $\sim$1/25. The bottom panel of Fig. 3 shows the variation in the angular separation between UVIT and Gaia sources as a function of distance from the center of SMC-1. The angular separation does not show any variation with respect to distance from center. Similar trend is also seen in SMC-2 and SMC-3.

\begin{deluxetable*}{ccccccccccc}
\tablecaption{Log of observations.\label{table-1}}
\tablehead{
\colhead{}         & \multicolumn{2}{c}{Field center}    & \colhead{} & \colhead{}  & \multicolumn{6}{c}{Exposure time in seconds for different filters} \\
\colhead{Field} & \colhead{$\alpha_{2000}$}  & \colhead{$\delta_{2000}$} & \colhead{Date} & \colhead{F154W} &  \colhead{F169M} & \colhead{F172M} &  \colhead{N245M} & \colhead{N263M} & \colhead{N279N} & \colhead{N219M} 
}
\startdata
SMC-1   & 01:09:46.0  & $-$71:20:30.0 & 31-12-2017 &  1995 & 2825 & 4982 &   2010  &  2011 & 3017 & 2878 \\
SMC-2   & 01:08:26.0  & $-$71:20:30.0 & 01-01-2018 &  2004 & 2953 & 5019 &   2028  &  2067 & 2978 & 2996 \\
SMC-3   & 01:09:46.0  & $-$71:26:30.0 & 01-01-2018 &  1993 & 2432 & 4810 &   2009  &  2011 & 2854 & 2425 \\ 
\enddata
\end{deluxetable*}

\section{Photometric Analysis} \label{sec:analysis}
The final combined and astrometric corrected images were analysed to get counts per second  for individual sources and these were then converted to the AB magnitudes as per the calibration given in \cite{2020AJ....159..158T}. There are two steps involved in making the best estimate of counts per second for individual sources. These steps are (i) an estimation based on a fit to a standard PSF to a small central part of the sources covering a radius of 3 sub-pixels for 
NUV and 4 sub-pixels for FUV. This was done to minimise any overlap with the neighboring sources in this crowded field and (ii) application of a correction factor to this counts per second,  to get the actual total counts per second in the full PSF. However, there is a small complication in this step. In the photon counting process used for UVIT, if multiple photons fall at the same location in any frame these are detected as a single photon. As the frame read rate is $\sim$29/s in full frame mode, the observed counts for a point source having 1 count per second  would suffer a saturation of $\sim$1.5\%, and the saturation would increase with increasing rate of counts. Thus, the correction factor for getting the actual total counts per second  involves a correction for saturation too.

\subsection{PSF photometry}
The procedure that was followed consists of (i) finding point sources in the field, (ii) modelling the PSF and (iii) fitting the PSF model to each of the detected point sources to obtain the instrumental magnitude. This procedure was carried out using the {\sc DAOPHOT} routines \citep{1987PASP...99..191S} implemented within {\sc IRAF}. Firstly, we detected all point sources using {\it daofind}  in each of the images based on the threshold, $N \times \sigma_{back}$. Here $\sigma_{back}$ is the standard deviation of the local background in the field and N is the threshold. We set N = 3 for all the images. However, this resulted in many incorrect detection of faint sources. So we smoothened the images by convolving them with a Gaussian with a $\sigma$ of 1.5 sub-pixels (0.62$''$).  which lead to improved source extraction. This improvement in detection after convolving with a Gaussian has also been noticed by \cite{2020ApJS..247...47L} on their analysis of M31 images from UVIT. Once the sources were detected on the smoothed images through the {\it daofind} task in IRAF which uses the centroiding algorithm, photometry was performed on the original un-smoothed images using the positions of the point sources returned by {\it daofind} on the smoothed images. To model the PSF, among the detected point sources, we selected about 10 relatively isolated stars in each of the SMC fields. The PSF model generated using those 10 stars was fit to each of the point sources found by {\it daofind} to get the instrumental magnitudes and the associated errors in them. They were then converted to AB magnitudes using the zero-point magnitudes given in \cite{2020AJ....159..158T}, and the errors in the AB magnitudes were obtained by error propagation \citep{1992drea.book.....B}. Various functional forms were used to model the PSF in IRAF such as gauss (elliptical gaussian function), lorentz (elliptical lorentzian function), moffat15 (elliptical Moffat function with a beta parameter of 1.5) and moffat25 (elliptical Moffat function with a beta parameter of 2.5), however, for generation of the final catalogue we adopted the PSF modelled using the moffat25 function, since moffat25 gave minimum residuals while modelling the PSF compared to other functions.

\begin{figure}
\vbox{
\hspace*{-0.8cm}\includegraphics[scale=0.2]{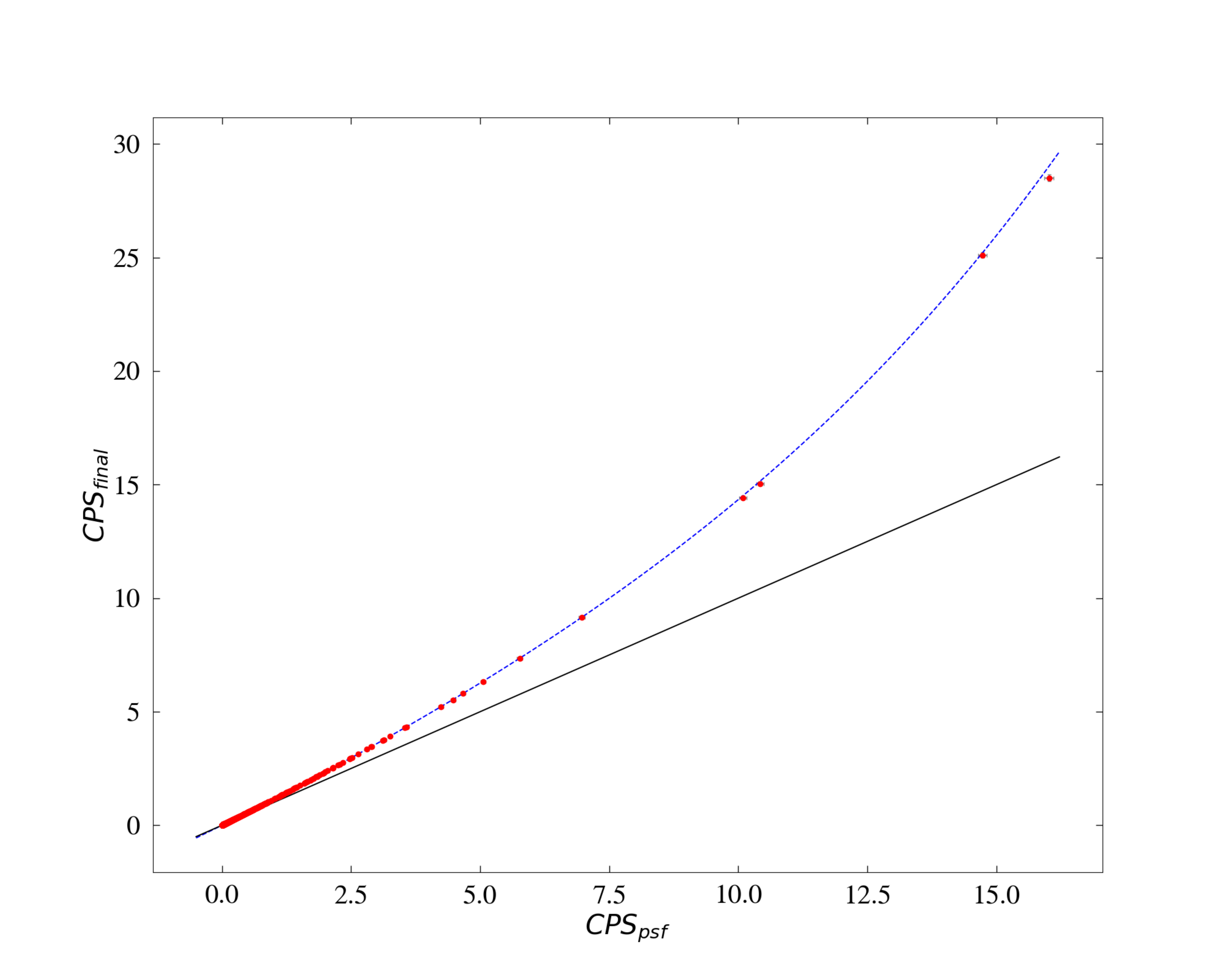}
\hspace*{-0.8cm}\includegraphics[scale=0.2]{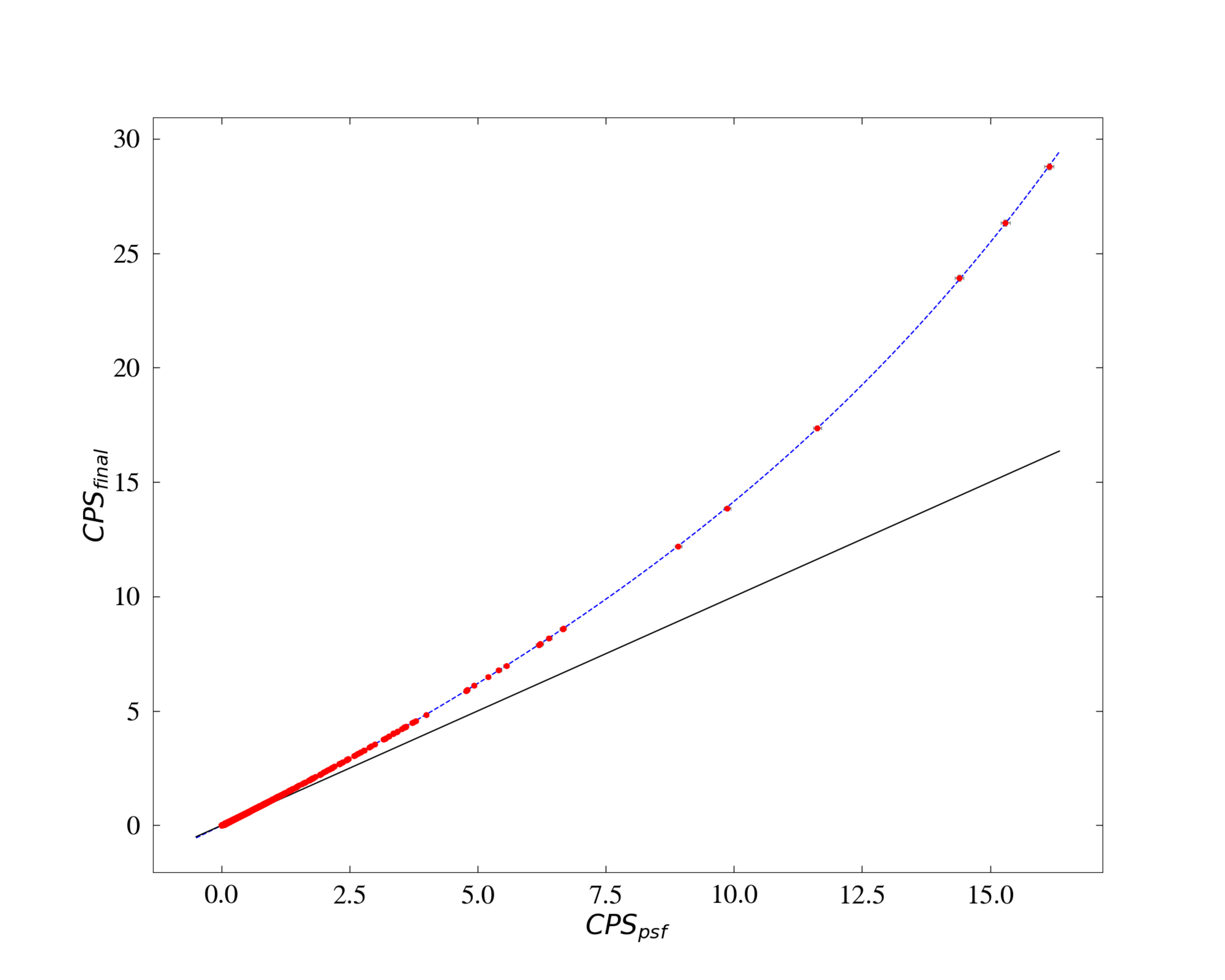}
     }
\caption{The correlation between the observed CPS from PSF fitting and the final corrected CPS for the FUV filter F154W (top panel) and the NUV filter N263M (bottom panel). The black solid line shows the $CPS_{psf}$ = $CPS_{final}$ line, while the blue dashed line is the empirical model in Equation 2.} 
\label{figure-empiricalmodel}
\end{figure}

\begin{figure*}
\hspace*{-0.8cm}\includegraphics[scale=0.25]{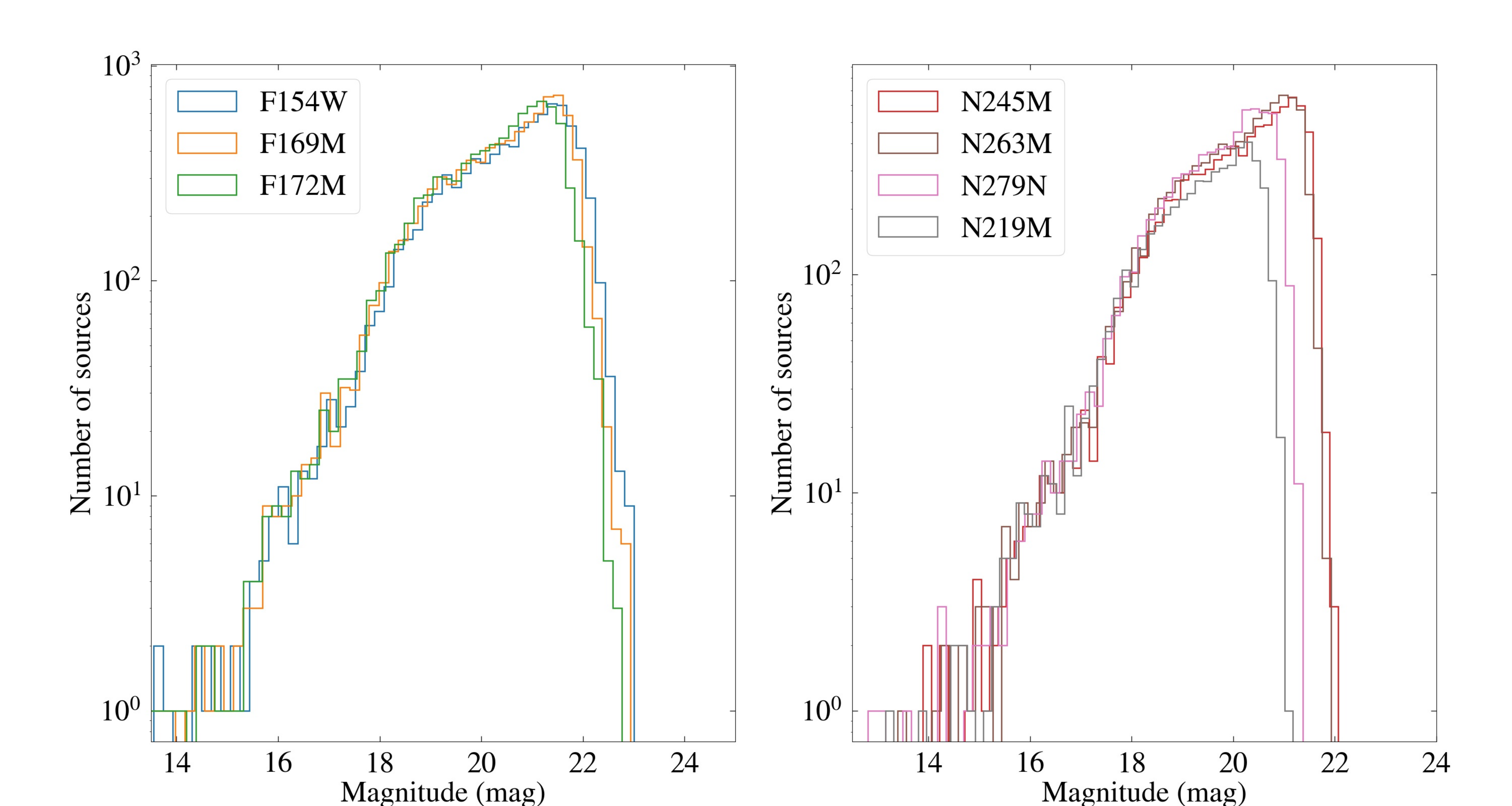}
\caption{Distribution of AB magnitudes of the point sources in the SMC field 
for the three FUV  (left panel) and four NUV (right panel) filters.} 
\label{figure-7}
\end{figure*}

\begin{figure*}
\centering
\gridline{\fig{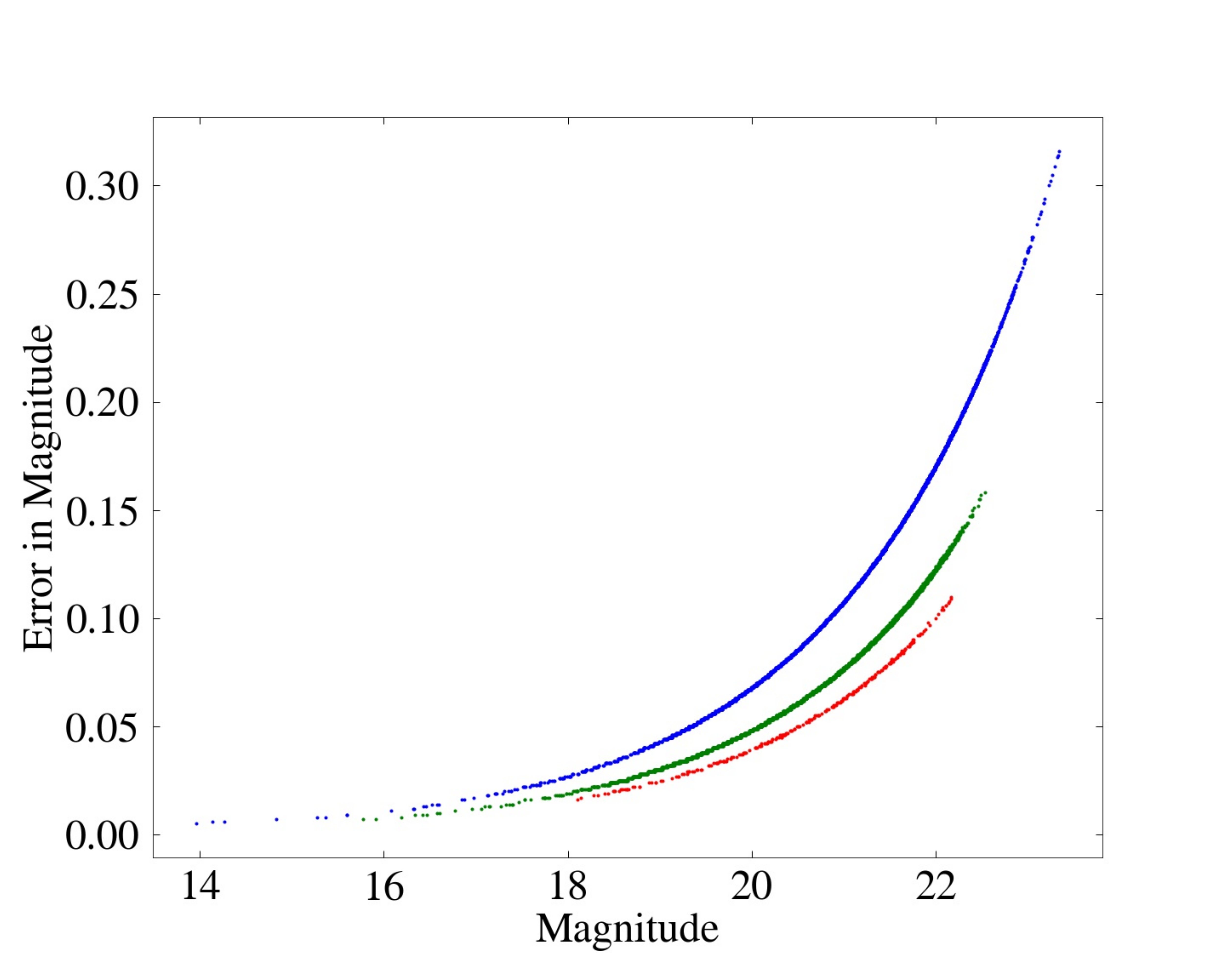}{0.34\textwidth}{(a) F154W}
          \fig{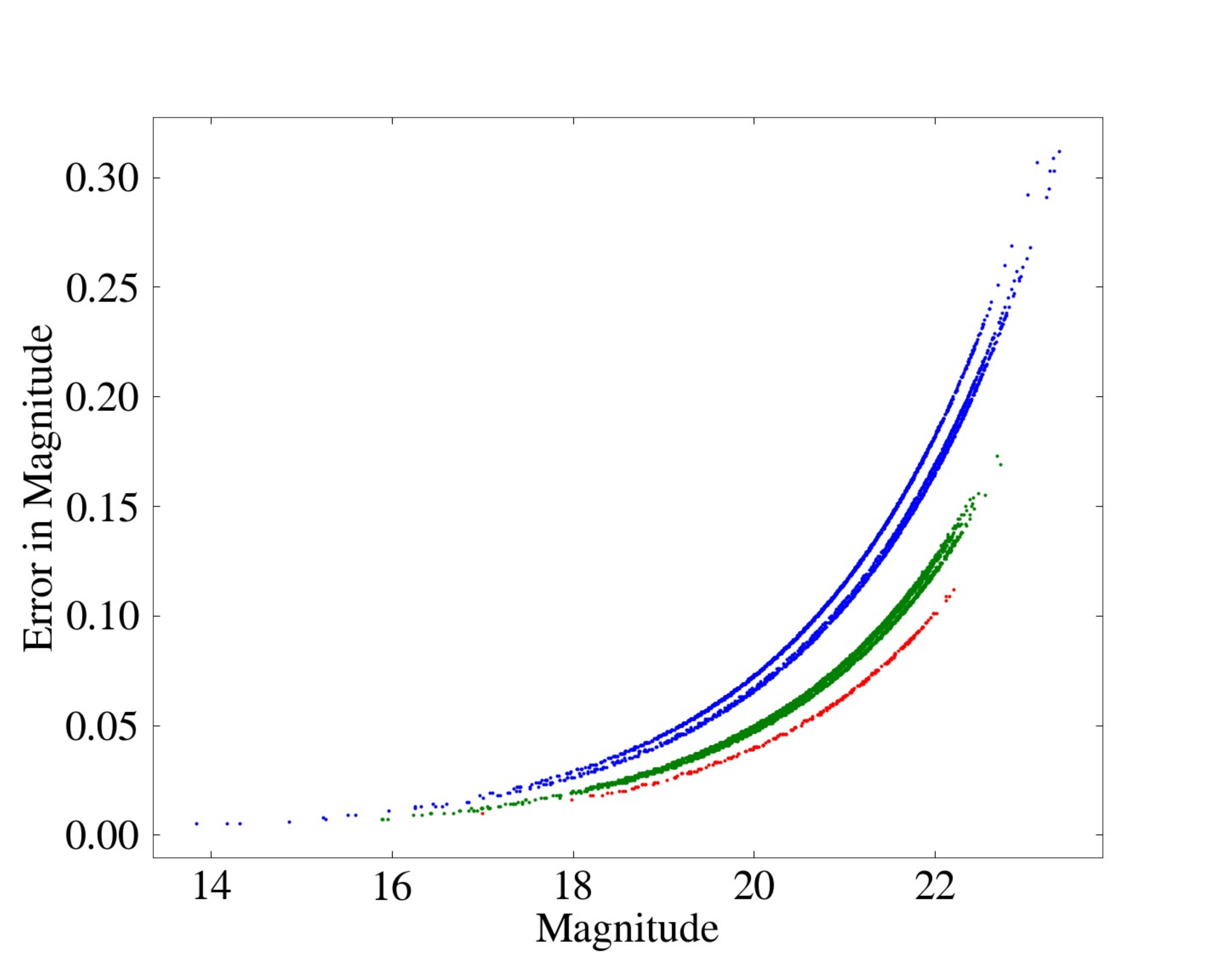}{0.34\textwidth}{(b) F169M}
          \fig{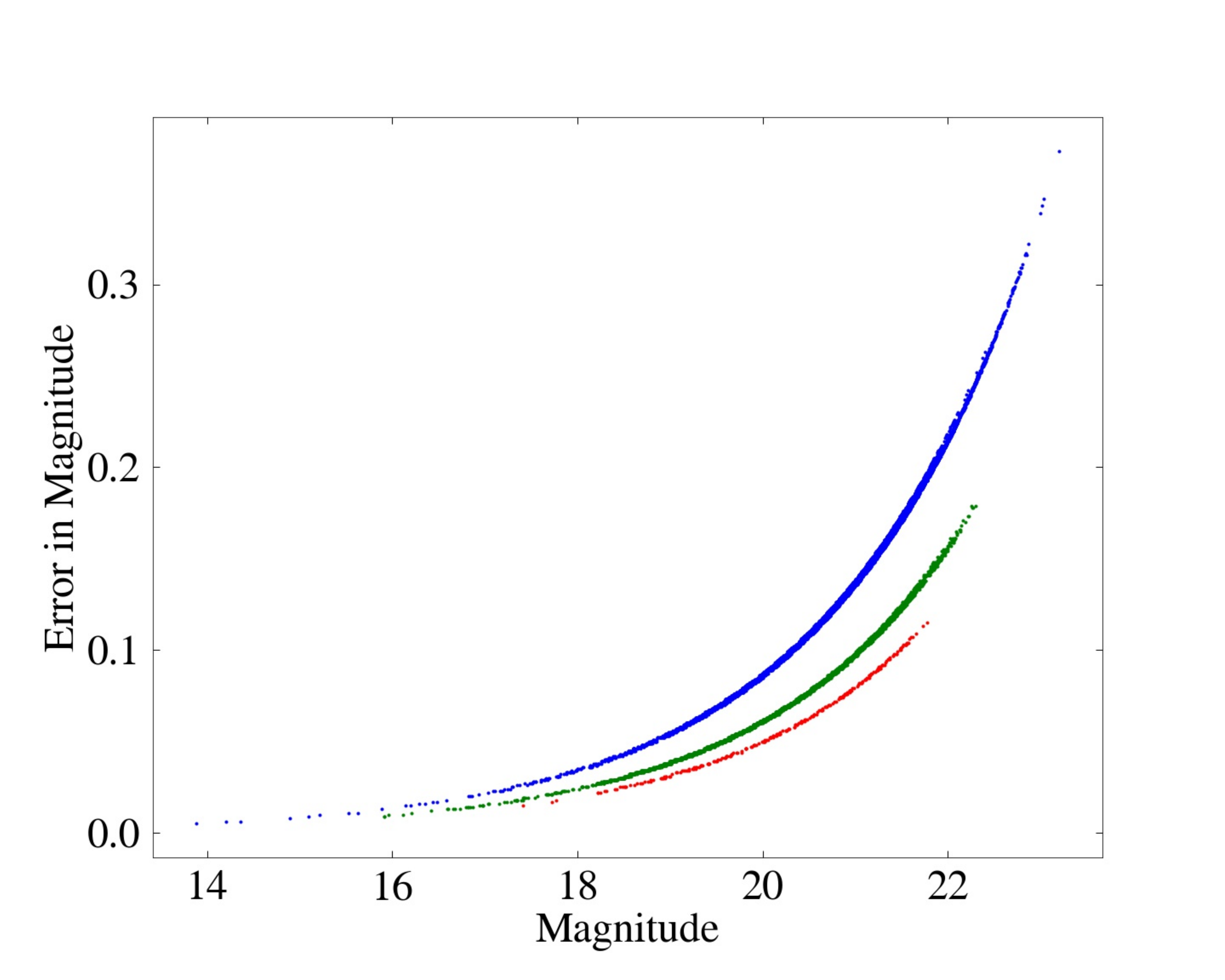}{0.34\textwidth}{(c) F172M}}
\gridline{\fig{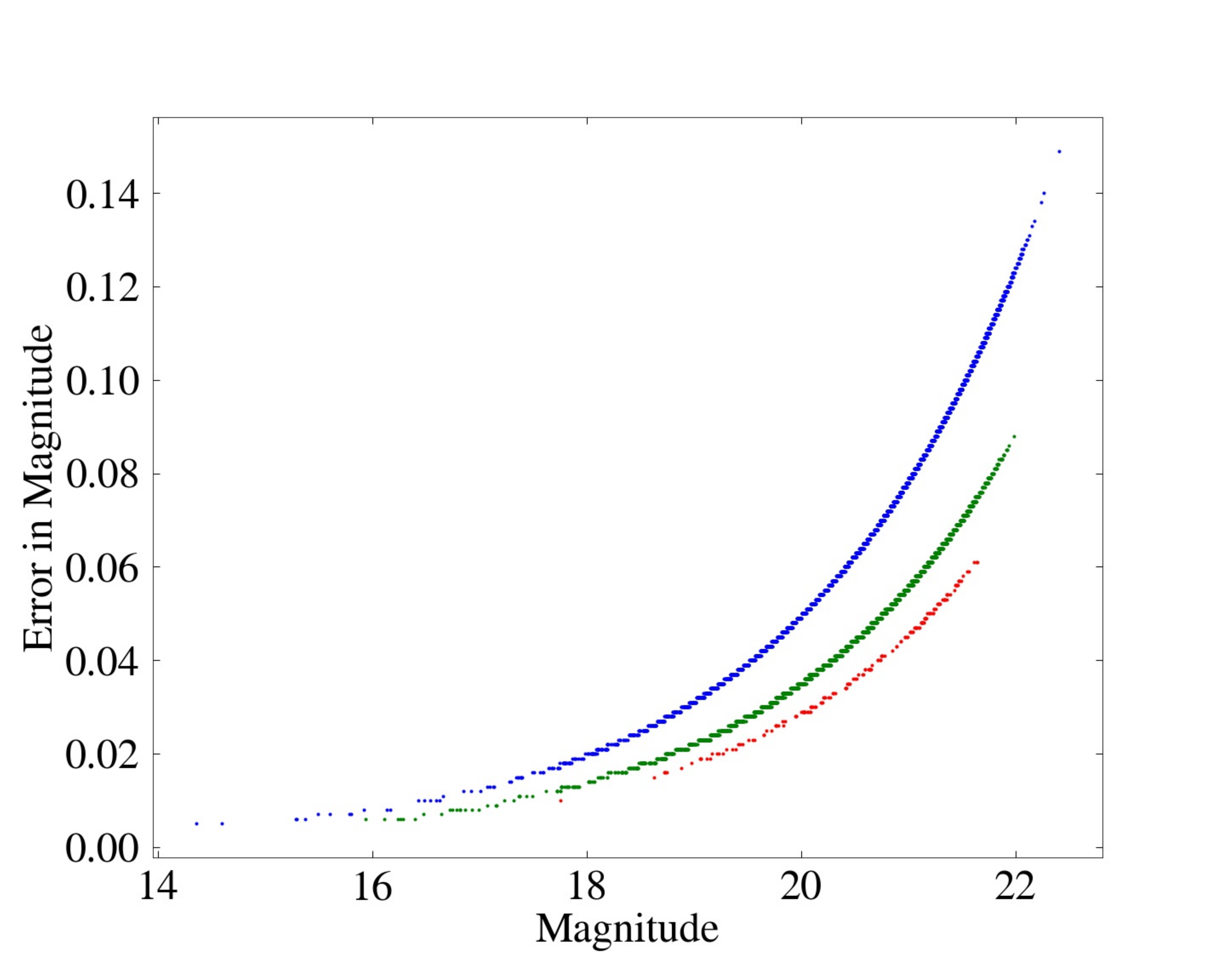}{0.34\textwidth}{(d) N245M}
          \fig{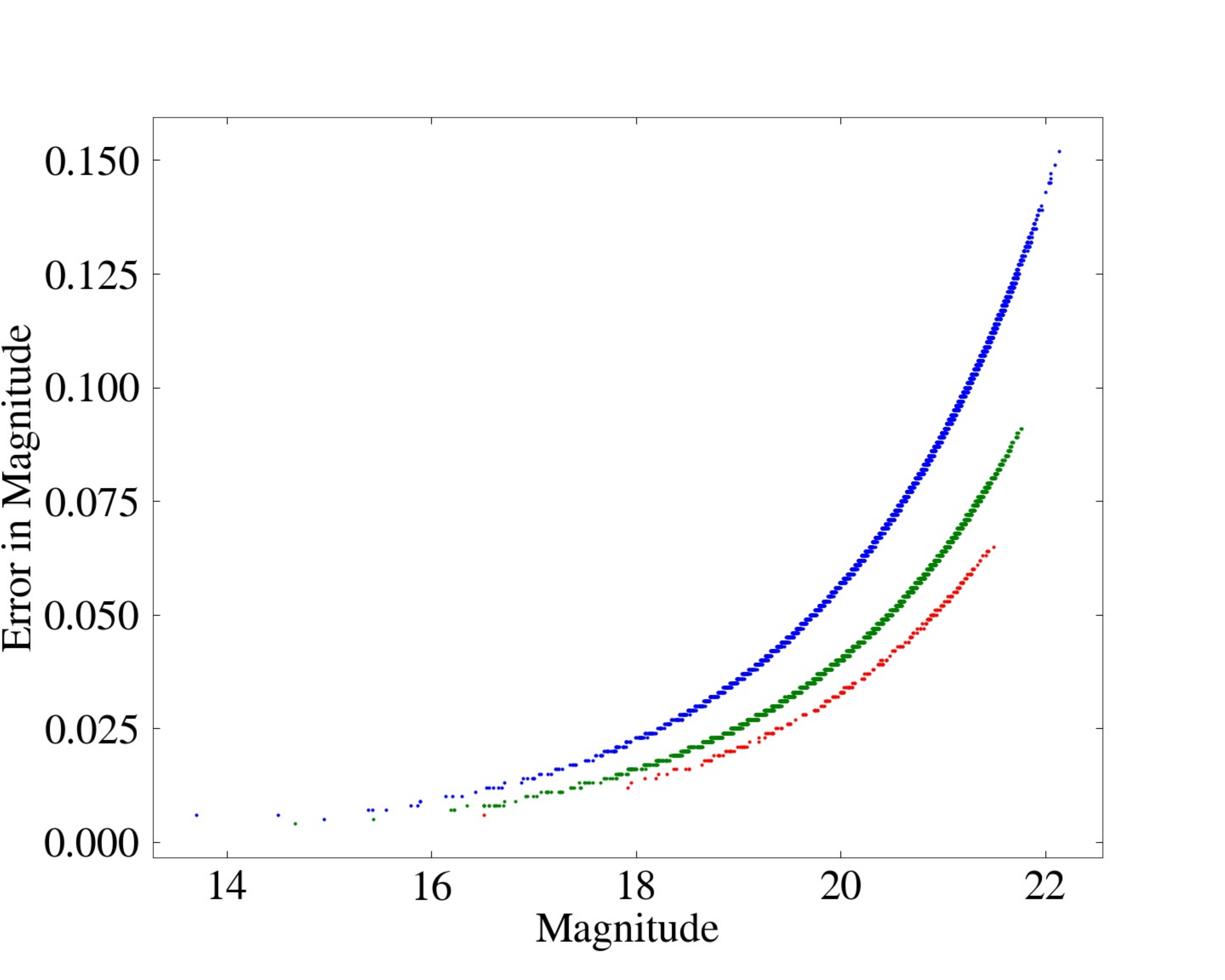}{0.34\textwidth}{(e) N263M}
          \fig{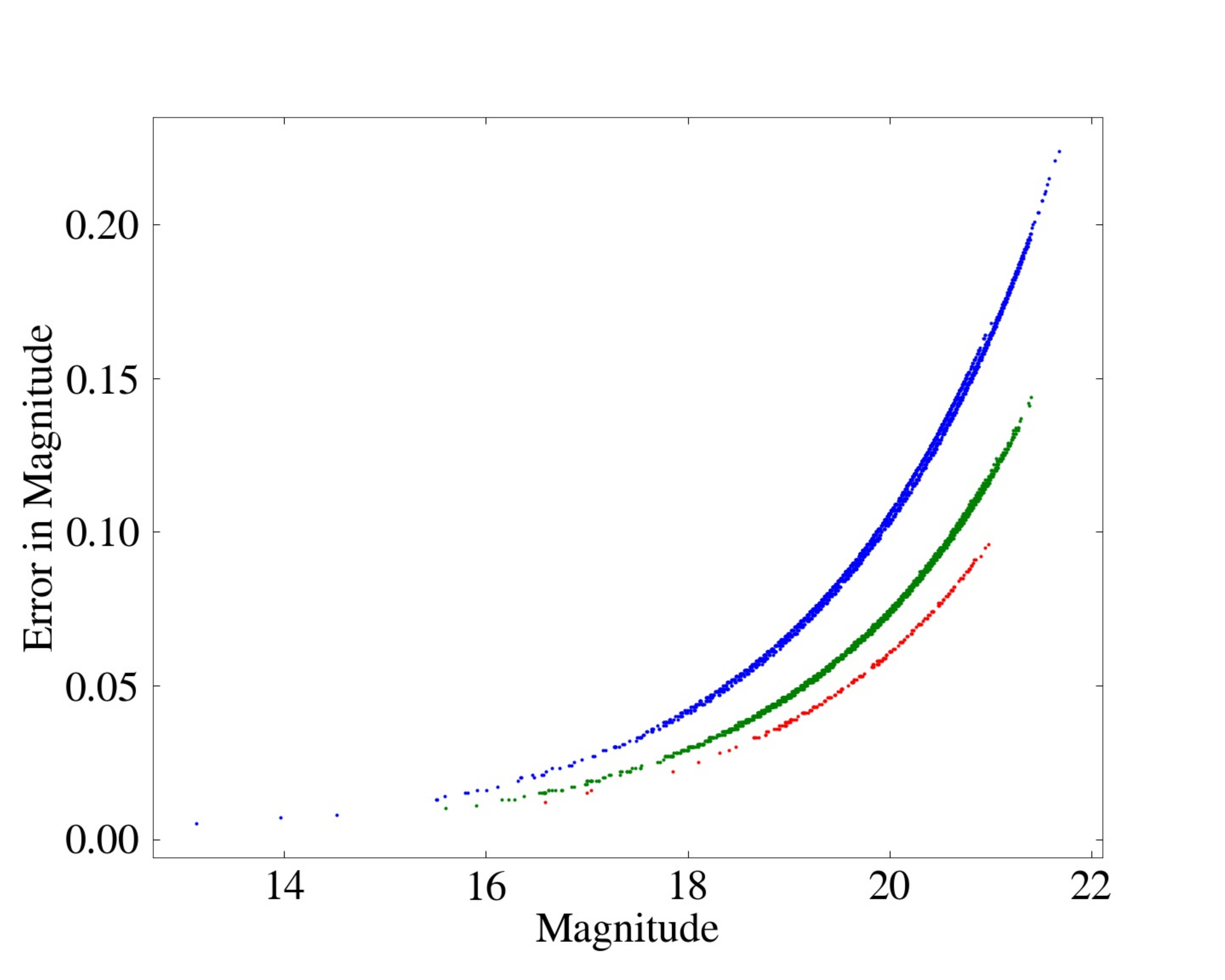}{0.34\textwidth}{(f) N279N}}
\gridline{\leftfig{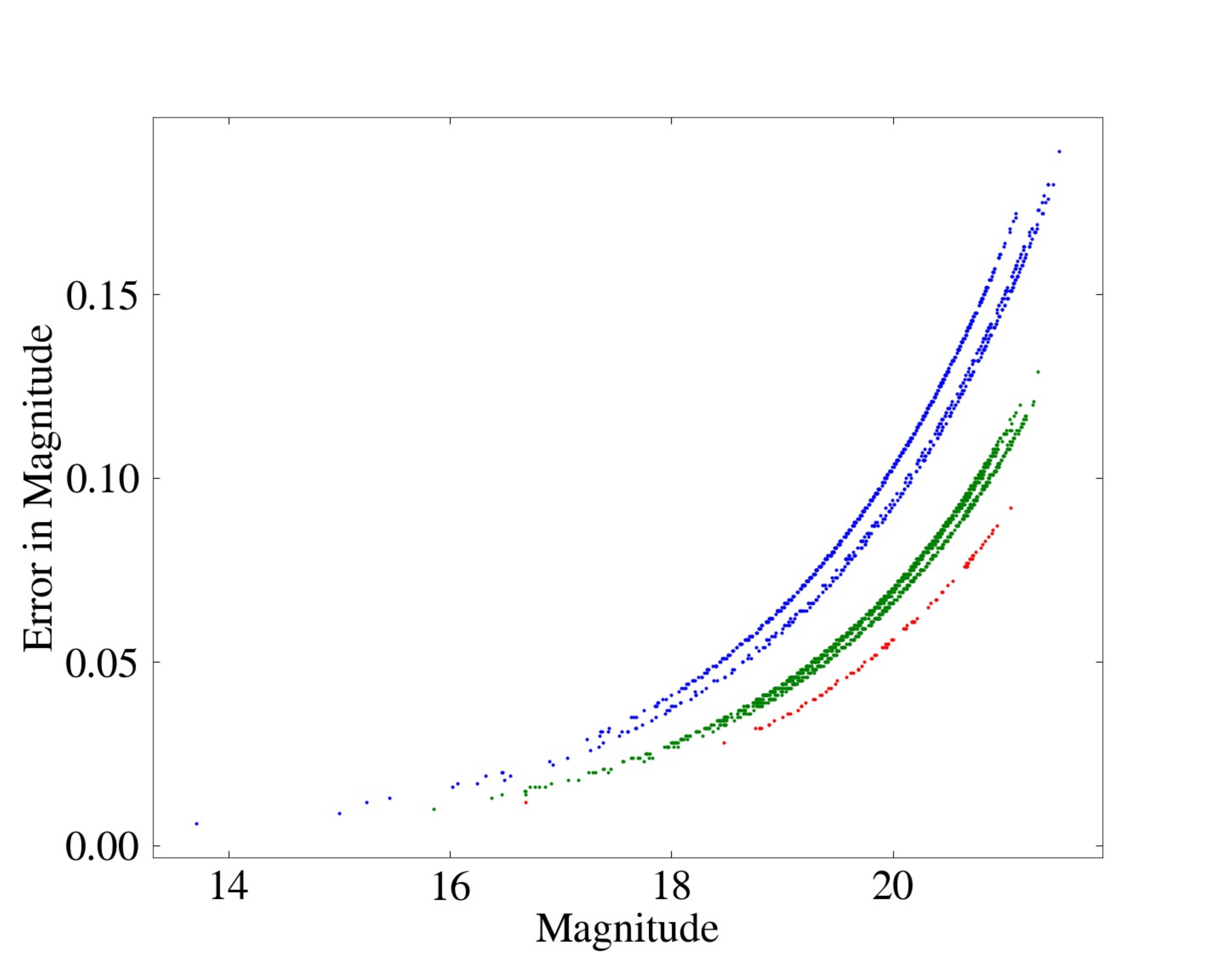}{0.34\textwidth}{(g) N219M}}
\caption{Error-magnitude plots for the detected sources in different filters. The three distinct bands in each filter correspond to sources common to all the three pointings (red), common in two pointings (green) and present in each individual pointing (blue). The splitting seen in blue and green are due to difference in exposure time between three different pointings (see Table \ref{table-1}).}
\label{figure-8}
\end{figure*}

\begin{deluxetable}{cccc}
\tablecaption{Details of the filters used for the observations.\label{table-2}}
\tablehead{
\colhead{Filter}      & \colhead{$\lambda_{mean} (\AA)$}    & \colhead{$\Delta \lambda (\AA)$}    & \colhead{Zero point magnitude}    
}
\startdata
F154W  & 1541  & 380 & 17.771 $\pm$ 0.01 \\
F169M  & 1608  & 290 & 17.410 $\pm$ 0.01 \\
F172M  & 1717  & 125 & 16.274 $\pm$ 0.02 \\
N245M  & 2447  & 280 & 18.452 $\pm$ 0.01 \\
N263M  & 2632  & 275 & 18.146 $\pm$ 0.01 \\
N279N  & 2792  & 90  & 16.416 $\pm$ 0.01 \\
N219M  & 2196  & 270 & 16.654 $\pm$ 0.02 \\
\enddata
\end{deluxetable}

\subsection{Estimation of total count rate in the full PSF of UVIT}
The counts obtained from PSF fit to the point sources need to be corrected for the counts in the outer part of the PSF and for saturation. As correction for saturation is a bit involved, we first describe the correction for counts in the outer part of the PSF while assuming that there is no 
saturation.  If there were isolated bright stars in the field, one could just find counts in a large aperture, e.g. in a radius of 30 sub-pixels which includes 97\% of the total counts (see \citealt{2020AJ....159..158T}). 
However, in this crowded field suitable isolated 
stars are not available, and we took a two step approach for this correction. 
Firstly, we found a conversion factor for the ratio of PSF-fitted flux to the flux 
in a radius of 12 sub-pixels using a selection of bright stars, and secondly used 
the conversion factor given in \cite{2020AJ....159..158T} to convert the 
flux in a radius of 12 sub-pixels to the total flux 
in the full PSF (100 sub-pixels radius).  The rationale
for choosing the intermediate step of finding the relative flux in a radius of 12 
sub-pixels is as follows: the core of the PSF, to which the PSF-fit is made, can change from 
image to image due to variations in errors of tracking the pointing and focus for 
individual filters, but the outer parts of the PSF are not affected by these 
small errors and thus the fractional energy content within a radius of 
12 sub-pixels is robust at $\sim$89\% (see \citealt{2020AJ....159..158T}).

Before explaining the various factors involved in correcting for saturation,  
let us get a rough idea of its magnitude. A rough estimate of the saturation 
can be made by invoking Poisson statistics for the total counts per frame, which 
is equal to counts per second  divided by 28.7 (the number of frames per second for the 
present observations). 
As the observed counts per frame is equal to “1 $-$ fraction of frames with no 
event/count”,  the saturation can be estimated from the following equation:
\begin{equation}
C = - ln (1 – F)
\end{equation}

where C is the corrected total counts per frame and F is the fraction of 
frames with no event/count.  However, the actual correction for saturation is less in the pedestal because the photons 
falling in the much less dense pedestal suffer very little saturation. To proceed 
further we followed the procedure described in \cite{2020AJ....159..158T}.
First, we assumed that all the saturation is limited within a radius of 
12 sub-pixels and that there is no saturation in the outer parts of the PSF. 
Next, we assumed that the saturation factor is constant within the radius of 
12 sub-pixels or the conversion factor from the PSF fitted counts per second  
to the counts per second  in the 
radius of 12 sub-pixels is unaffected by saturation. Given these two assumptions, for every value of PSF fitted counts per second, the saturation corrected 
total counts per second  can be calculated from the equations for saturation and the detailed PSF given in \cite{2020AJ....159..158T}. We found that the PSF fitted counts per second  and the total corrected counts per second  are well fitted by the equation

\begin{equation}
\small
CPS_{final}  = X \times \left(\frac{1}{CF_{12}} + SAT (X)\right)
\end{equation}

Here, $CPS_{final}$ is the final corrected counts per second for the full PSF, X = CPS$_{PSF}$ $\times$ CF$_{PSF12}$, where CPS$_{PSF}$ is the counts per second in the fitted PSF, CF$_{PSF12}$ is the correction factor for correcting the PSF fitted counts per second to counts per second in a radius of 12 sub-pixels. The first term on the right hand side of Equation 2, gives the total counts per second  without any correction for saturation, and the second term represents the correction for saturation. Values for the conversion factor CF$_{PSF12}$ for the various filters are given in Table \ref{table-aperturecorrection}, while CF$_{12}$, the correction factor to covert the counts per second  from 12 sub-pixels radius to counts per second in the full PSF (100 sub-pixels radius) is 0.893 for FUV and 0.886 for
NUV (from \citealt{2020AJ....159..158T}). Values for the function SAT(X) are well fitted by the following polynomial of third order as

\begin{equation}
SAT (X)  =  a1 + a2 \times X + a3 \times X^2 + a4 \times X^3
 \label{Correction_polynomial_saturation}
\end{equation}

The coefficients of this polynomial for NUV and FUV are given in Table 3.
Details of the procedure for the calculation of the function “SAT” 
are given in Appendix (also see Fig. 4).

All the above discussion on saturation refers to the actual observed counts per second on the detector, while the counts per second  in the images involve a correction for the flat field.  Therefore, we first have to calculate the actual observed counts per second on the detector by applying the flat field correction in reverse, calculate the total counts per second  for this corrected rate and finally apply the flat-field correction to this corrected rate. The flat field correction factor used for this is an average of its values over 21 $\times$ 21 sub-pixels ($\sim$ 1.1$^{\prime}$ $\times$ 1.1$^{\prime}$) around the centre of the source to account for drift during the pointing. Finally, we note that this correction for saturation is accurate to 5\% for observed counts per second $<$ 12 within a radius of 12 sub-pixels. We also note that we have neglected another saturation effect which is related to the saturation current in the MCP of the detector. This depends on the counts per second,  and is estimated to be $<$ 5\% for 150 counts per second (see \citealt{2020AJ....159..158T}).

\begin{deluxetable}{ccc}
\tablecaption{Coefficients in Equation \ref{Correction_polynomial_saturation} \label{table-polynomial_coefficients}}.
\tablewidth{0pt}
\tablehead{
\colhead{Coefficient}  & \colhead{FUV}    & \colhead{NUV}   
}
\startdata
a1  & -0.003016 & -0.002775 \\
a2  & 0.024022  & 0.023266 \\
a3  & -0.000142  & -9.669652 $\times$ $10^{-5}$ \\
a4  & 8.215584 $\times$ $10^{-5}$ & 7.507352 $\times$ $10^{-5}$ \\
\enddata
\end{deluxetable}

\subsection{Completeness of the catalogue}
We show in Fig. \ref{figure-7} the magnitude distribution of the sources detected in the FUV and NUV filters. The peak of the magnitude distribution gives an approximate estimate of the completeness of the SMC observations. In the FUV band, for the filters F154W, F169M and F172M we found the peak in the distribution of magnitudes at 21.30, 21.41 and 21.09 mag respectively. Similarly, for the NUV channel we found values of 21.08, 20.89, 20.34 and 20.21 mag respectively for the filters N245M, N263M, N279N and N219M. The variation of error as a function of brightness for all the filters are shown in Fig. \ref{figure-8}. The errors show sharp increasing trend after magnitudes that roughly correspond to the peak of the distribution in Fig. \ref{figure-7}.

\begin{figure}
\centering
\vbox{
\hspace*{-0.8cm}\includegraphics[scale=0.2]{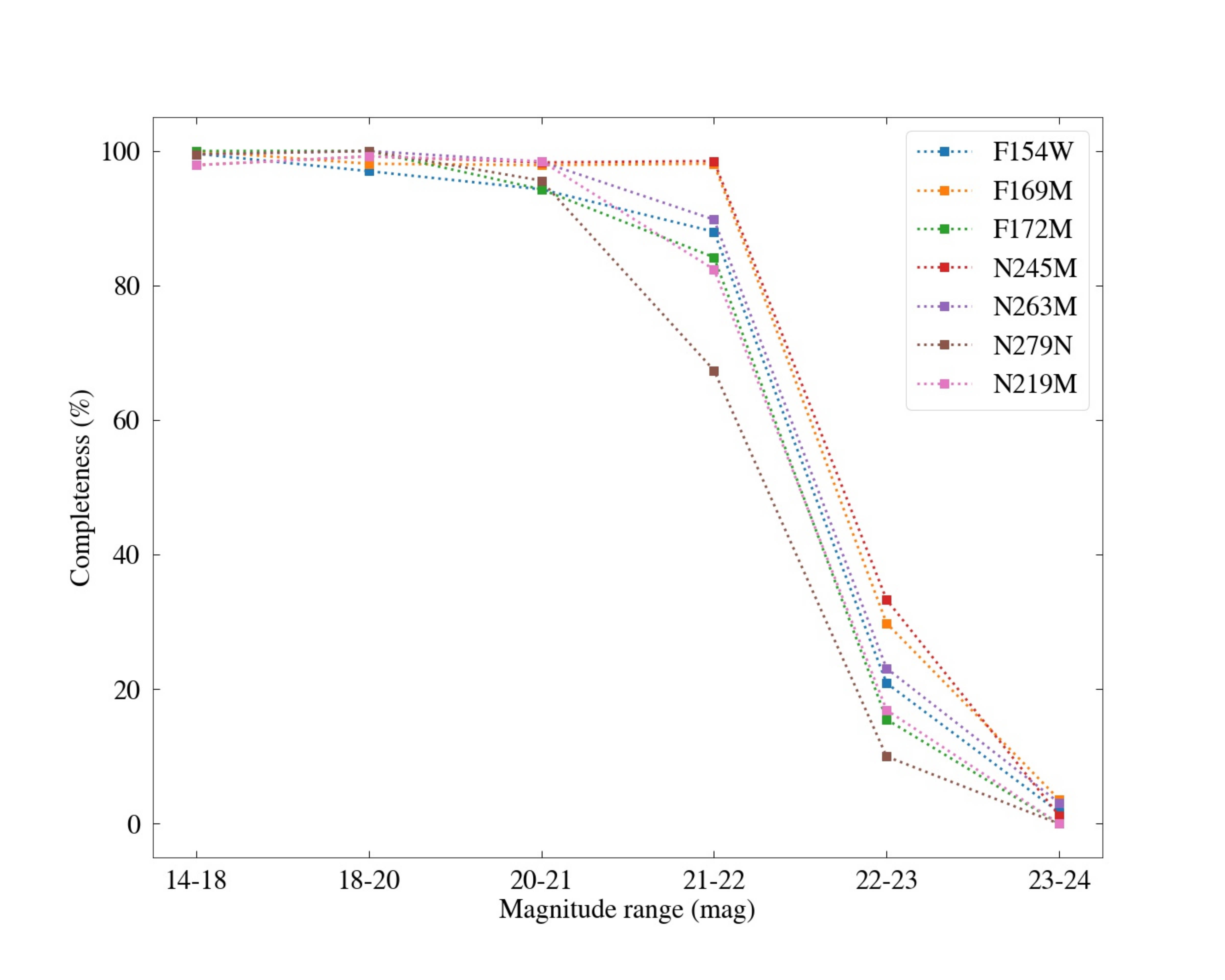}
     }
\caption{The completeness in percentage for different filters in the catalogue.} 
\label{figure-completeness}
\end{figure}

We also assessed the completeness of our photometry as a function of brightness by introducing artificial stars. We added artificial stars numbering about 10\% of the point sources detected in each of the filters. They with pre-selected positions and brightness were added randomly (using the {\it addstar} routine in {\sc IRAF}) to each of the filters, so as not to alter the crowding characteristics. After the addition of the artificial stars, the photometry of the frames were carried out in the usual procedure (see Section 3.1). The ratio of the number of recovered stars to that inserted gives a measure of the completeness of our photometry. The variation of the completeness factor as a function of brightness for different filters is given in Table \ref{table-completeness} and shown in Fig. \ref{figure-completeness}.

\begin{deluxetable*}{ccrrrrrrrr}
\tablecaption{Aperture correction and flux ratio (based on PSF magnitudes and the magnitudes obtained over a radius of 12 sub-pixels) in different NUV and FUV filters for different PSF fitting models. \label{table-aperturecorrection}}
\tablewidth{0pt}
\tabletypesize{\scriptsize}
\tablehead{
\colhead{Filter}  & \colhead{No.of}    & \multicolumn{2}{c}{gauss}  & \multicolumn{2}{c}{moffat25} & \multicolumn{2}{c}{moffat15} & \multicolumn{2}{c}{lorentz}  \\
\colhead{} & \colhead{stars}  & \colhead{AC} & \colhead{Ratio} & \colhead{AC}  & \colhead{Ratio} & \colhead{AC} & \colhead{Ratio}  & \colhead{AC} & \colhead{Ratio} 
}
\startdata
F154W & 81 & -0.276 $\pm$ 0.004 & 1.290 $\pm$ 0.004 & -0.299 $\pm$ 0.004 & 1.318 $\pm$ 0.005 & -0.312 $\pm$ 0.004 & 1.333 $\pm$ 0.005 & -0.335 $\pm$ 0.004 & 1.362 $\pm$ 0.004 \\
F169M & 109 & -0.322 $\pm$ 0.003 & 1.346 $\pm$ 0.004 & -0.335 $\pm$ 0.003 & 1.362 $\pm$ 0.004 & -0.340 $\pm$ 0.003 & 1.369 $\pm$ 0.004 & -0.349 $\pm$ 0.003 & 1.380 $\pm$ 0.004 \\
F172M & 72 & -0.298 $\pm$ 0.004 & 1.317 $\pm$ 0.005 & -0.311 $\pm$ 0.004 & 1.332 $\pm$ 0.005 & -0.319 $\pm$ 0.004 & 1.342 $\pm$ 0.005 & -0.336 $\pm$ 0.004 & 1.363 $\pm$ 0.005 \\
N245M & 114 & -0.415 $\pm$ 0.004 & 1.467 $\pm$ 0.006 & -0.417 $\pm$ 0.004 & 1.469 $\pm$ 0.006 & -0.416 $\pm$ 0.004 & 1.468 $\pm$ 0.006 & -0.429 $\pm$ 0.004 & 1.486 $\pm$ 0.006 \\
N263M & 110 & -0.413 $\pm$ 0.005 & 1.465 $\pm$ 0.007 & -0.416 $\pm$ 0.005 & 1.469 $\pm$ 0.007 & -0.412 $\pm$ 0.005 & 1.463 $\pm$ 0.007 & -0.434 $\pm$ 0.005 & 1.493 $\pm$ 0.007 \\
N279N & 49 & -0.397 $\pm$ 0.007 & 1.442 $\pm$ 0.009 & -0.405 $\pm$ 0.007 & 1.454 $\pm$ 0.009 & -0.400 $\pm$ 0.007 & 1.447 $\pm$ 0.009 & -0.432 $\pm$ 0.007 & 1.491 $\pm$ 0.009 \\
N219M & 75 & -0.528 $\pm$ 0.007 & 1.629 $\pm$ 0.011 & -0.531 $\pm$ 0.007 & 1.633 $\pm$ 0.011 & -0.533 $\pm$ 0.007 & 1.636 $\pm$ 0.010 & -0.544 $\pm$ 0.006 & 1.652 $\pm$ 0.010 \\ \hline
\enddata
\end{deluxetable*}

\begin{deluxetable*}{cccccccc}
\tablecaption{Variation of completeness of the catalogue with the brightness of the sources 
\label{table-completeness}}
\tablewidth{0pt}
\tablehead{
\colhead{Mag. range}  & \colhead{F154W}    & \colhead{F169M}  & \colhead{F172M} & \colhead{N245M} & \colhead{N263M} & \colhead{N279N} & \colhead{N219M}}
\startdata
14-18 & 1.00  & 1.00 & 1.00 & 0.98 & 1.00 & 1.00 & 0.98 \\
18-20 & 0.97  & 0.98 & 1.00 & 0.99 & 1.00 & 1.00 & 0.99 \\
20-21 & 0.94  & 0.98 & 0.94 & 0.98 & 0.98 & 0.96 & 0.98 \\
21-22 & 0.88  & 0.98 & 0.84 & 0.99 & 0.90 & 0.67 & 0.82 \\
22-23 & 0.21  & 0.30 & 0.16 & 0.33 & 0.23 & 0.10 & 0.17 \\
23-24 & 0.02  & 0.04 & 0.00 & 0.01 & 0.03 & 0.00 & 0.00 \\ \hline
\enddata
\end{deluxetable*}

\begin{deluxetable*}{ccccccccccccccccc}
\tablecaption{A sample of 15 sources from the matched catalogue of point sources in SMC in three FUV and four NUV filters \label{table-samplecatalogue}}
\tablewidth{0pt}
\tabletypesize{\scriptsize}
\tablehead{
\colhead{S. No.}  & \colhead{$\alpha_{2000}$ (deg)}    & \colhead{$\delta_{2000}$(deg)}  & \multicolumn{2}{c}{F154W}  & \multicolumn{2}{c}{F169M} & \multicolumn{2}{c}{F172M} & \multicolumn{2}{c}{N245M} & \multicolumn{2}{c}{N263M} & \multicolumn{2}{c}{N279N} & \multicolumn{2}{c}{N219N} \\
\colhead{} & \colhead{}  & \colhead{} & \colhead{AB} & \colhead{Error} & \colhead{AB}  & \colhead{Error} & \colhead{AB} & \colhead{Error} & \colhead{AB} & \colhead{Error} & \colhead{AB} & \colhead{Error} & \colhead{AB} & \colhead{Error} & \colhead{AB} & \colhead{Error} 
}
\startdata
1 & 17.483551 & -71.477719 & 19.274 & 0.049 & 19.206 & 0.047 & 19.194 & 0.059 & 19.127 & 0.033 & 19.034 & 0.036 & 18.999 & 0.065 & 18.992 & 0.059 \\
2 & 17.501380 & -71.474138 & 19.861 & 0.064 & 19.638 & 0.057 & 19.817 & 0.079 & 19.870 & 0.047 & 19.655 & 0.049 & 19.612 & 0.086 & 19.354 & 0.070 \\
3 & 17.411340 & -71.473069 & 21.085 & 0.112 & 21.079 & 0.111 & 20.556 & 0.111 & 20.625 & 0.066 & 20.402 & 0.068 & 20.453 & 0.127 & 20.244 & 0.106 \\
4 & 17.516082 & -71.469507 & 16.307 & 0.012 & 16.281 & 0.012 & 16.355 & 0.016 & 16.361 & 0.009 & 16.322 & 0.010 & 16.377 & 0.019 & 16.217 & 0.017 \\
5 & 17.426452 & -71.468917 & 20.073 & 0.070 & 20.053 & 0.069 & 19.800 & 0.078 & 19.392 & 0.037 & 19.257 & 0.040 & 19.223 & 0.072 & 19.383 & 0.071 \\
6 & 17.546273 & -71.468802 & 20.399 & 0.082 & 20.105 & 0.071 & 20.354 & 0.101 & 20.214 & 0.055 & 20.149 & 0.061 & 20.020 & 0.104 & 20.007 & 0.095 \\
7 & 17.442353 & -71.467313 & 20.776 & 0.097 & 20.816 & 0.098 & 20.722 & 0.119 & 20.704 & 0.068 & 20.415 & 0.069 & 20.343 & 0.121 & 20.348 & 0.111 \\
8 & 17.218003 & -71.463924 & 21.673 & 0.147 & 21.519 & 0.135 & 21.242 & 0.152 & 20.926 & 0.076 & 20.469 & 0.071 & 19.995 & 0.103 & 20.581 & 0.123 \\
9 & 17.332593 & -71.463868 & 19.558 & 0.055 & 19.511 & 0.054 & 19.564 & 0.070 & 19.589 & 0.041 & 19.494 & 0.045 & 19.235 & 0.072 & 19.401 & 0.072 \\
10 & 17.548006 & -71.462470 & 19.460 & 0.053 & 19.380 & 0.051 & 19.239 & 0.060 & 19.085 & 0.032 & 18.945 & 0.035 & 18.877 & 0.061 & 19.004 & 0.060 \\
11 & 17.657541 & -71.459780 & 21.125 & 0.114 & 21.621 & 0.142 & 21.294 & 0.155 & 21.068 & 0.081 & 20.820 & 0.083 & 20.774 & 0.147 & 20.699 & 0.130 \\
12 & 17.542935 & -71.458877 & 21.101 & 0.113 & 21.374 & 0.127 & 20.951 & 0.133 & 20.993 & 0.078 & 20.598 & 0.075 & 20.195 & 0.113 & 20.533 & 0.121 \\
13 & 17.443856 & -71.458562 & 21.740 & 0.151 & 21.481 & 0.133 & 21.444 & 0.166 & 21.265 & 0.088 & 21.284 & 0.103 & 20.749 & 0.145 & 20.726 & 0.132 \\
14 & 17.666746 & -71.456551 & 20.290 & 0.078 & 20.238 & 0.075 & 20.164 & 0.092 & 20.040 & 0.050 & 19.816 & 0.052 & 19.720 & 0.091 & 19.847 & 0.088 \\
15 & 17.537647 & -71.455490 & 20.427 & 0.083 & 20.316 & 0.078 & 20.184 & 0.093 & 20.267 & 0.056 & 19.953 & 0.056 & 19.913 & 0.099 & 20.157 & 0.102 \\
\enddata
\tablecomments{The table in full is available in the electronic version of the article}
\end{deluxetable*}

\section{Summary} \label{sec:summary}
In this work, we have analysed three pointings of SMC, observed by UVIT. From 
this analysis, we arrived at a catalogue of 11,241 UV sources in the three fields 
of SMC, and provide their AB magnitudes in a total of seven filters, three in 
FUV and four in  NUV. The sample catalogue of 15 sources is given in 
Table \ref{table-samplecatalogue}. The full catalogue is available in the 
electronic version of the article. This catalogue will be of use to the 
astronomical community to address a large range of astronomical problems. We 
also carried out an evaluation of the relation between observed and saturation 
corrected UVIT magnitudes. We found that the observed UVIT magnitudes need to 
be corrected for the effects of saturation and PSF and provide empirical 
relations for the same.

\acknowledgments
We thank the anonymous referee for his/her critical comments that helped to improve the manuscript. This publication uses the data from the AstroSat mission of the Indian Space 
Research Organisation (ISRO), archived at the Indian Space Science Data Centre 
(ISSDC). This publication uses UVIT data processed by the payload operations 
centre at IIA (Indian Institute of Astrophysics). The UVIT is built in 
collaboration between IIA, IUCAA (Inter University Center for Astronomy and Astrophysics), 
TIFR (Tata Institute of Fundamental Research), ISRO and 
CSA (Canadian Space Agency). 
\software{IRAF \citep{1986SPIE..627..733T}, Astropy \citep{2013A&A...558A..33A}, Scipy \citep{2020SciPy-NMeth}, Numpy \citep{harris2020array}
Pandas \citep{mckinney2010data}, Matplotlib \citep{Hunter:2007}
}

\appendix
\section{Saturation correction}
In the NUV and FUV channels of UVIT, the occurrence of multiple photon events within 3 $\times$ 3 pixels ($\sim 10''\times 10''$) in a frame is detected as a single photon. Thus, for point sources with counts per frame $>$ 0.1, significant number of photons are not recorded. This effect of saturation needs to be taken into account while estimating
the brightness of point sources from the measured count rates. Another possible
source of saturation is the reduced multiplication of the photo electrons in the
MCPs, reducing the final signal and hence reducing the probability of detection, when the local photon rate is high. However, this effect is estimated to be $<$ 5\% for rates of 150 detected photons per second for a point source and has been ignored \citep{2017AJ....154..128T}. To correct for the effects of saturation, we adopted the empirical method outlined in \cite{2017AJ....154..128T}. Essential steps in this method are: i) assuming Poisson statistics for the occurrence of multiple photons in a frame for 97\% of the counts per frame in the full PSF (CPF5), an ideal correction (ICORR) was calculated (97\% is arbitrarily chosen to discard counts in the outermost part of the PSF), ii) a relation (given in Equation \ref{eqn:Tandon_saturation}) was derived relating the ideal correction to the actual correction found from detailed analysis of the frames, and iii) the actual correction was used to find the corrected counts per frame and hence the corrected counts per second as given in Equation \ref{eqn:Tandon_finalcps}.

\begin{equation}\label{eqn:Tandon_saturation}
RCORR = ICORR ( 0.89 - 0.30 \times ICORR^2)
\end{equation}

\begin{equation}\label{eqn:Tandon_finalcps}
CPS = (CPF5/0.97 + RCORR) \times 28.7
\end{equation}

Where 28.7 is the number of frames per second for the present observations. As we used the observed counts in a radius of 12 sub-pixels, the above process needs to be translated to obtain the corrected total counts per second from those observed in a radius of 12 sub-pixels. To do this, for various values of the corrected counts per second within the full PSF (CPS), the expected counts per second within a radius of 12 sub-pixels (CO12) were calculated as per the prescription in Equation \ref{eqn:Tandon_saturation}, \ref{eqn:Tandon_finalcps} and \ref{eqn:cpsin12pix}. We also assumed that all the saturation is limited to a radius of 12 sub-pixels. The values of CPS were fitted to a third order polynomial in CO12. This process is illustrated in the following equations:

\begin{equation}\label{eqn:cpsin12pix}
CO12 = CPS \times CF_{12} - RCORR \times 28.7
\end{equation}

Here, $CF_{12}$ is the correction factor to convert the counts per second in the full PSF (100 sub-pixels radius) to those in a radius of 12 sub-pixels in the absence of saturation, which is 0.893 for NUV and 
0.886 for FUV \citep[see Table 11 of][]{2020AJ....159..158T}.

We define a saturation correction factor SAT as per Equation \ref{eqn:sat_definition} and calculated SAT and CO12 for various values of CPS in the range 0.6 to 17. Next a polynomial fit was made relating SAT to CO12 as

\begin{equation}\label{eqn:sat_definition}
CPS  = CO12 \times \left(\frac{1}{CF_{12}} + SAT \right)
\end{equation}

\begin{equation}
SAT(CO12) = a1 + a2 \times CO12 + a3 \times CO12^2 + a4 \times CO12^3
\label{correction_polynomial}
\end{equation}

The coefficients of Equation \ref{correction_polynomial} are given in Table \ref{table-polynomial_coefficients}.

\bibliography{ref}{}
\bibliographystyle{aasjournal}

\end{document}